# Partial identification of Mean Achievement in ILSA studies with Multi-Stage Stratified Sample Design and Student Non-Participation[1]

Diego Cortes

International Association for the Evaluation of Educational Achievement (Germany)

Jeff Dominitz

Department of Economics, Rice University (USA)

Maximiliano Romero

International Association for the Evaluation of Educational Achievement (Germany)

## Abstract

International large-scale assessment (ILSA) studies collect information across education systems with the objective of learning about the population-wide distribution of student achievement in the assessment. In this article, we study one of the most fundamental threats that these studies face when justifying the conclusions reached about these distributions: the identification problem that arises from student non-participation during data collection. Recognizing that ILSA studies have traditionally employed a narrow range of strategies to address non-participation, we examine this problem using tools developed within the framework of partial identification of probability distributions. We tailor this framework to the problem of non-participation when data are collected using a multi-stage stratified random sample design, as in most ILSA studies. We demonstrate this approach with application to the International Computer and Information Literacy Study in 2018. We show how to use the framework to assess mean achievement under reasonable and credible sets of assumptions about the non-participating population. We also provide examples of how these results may be reported by agencies that administer ILSA studies. By doing so, we bring to the field of ILSA an alternative strategy for identification, estimation, and reporting of population parameters of interest.

[1] Corresponding author: Diego Cortes (diego.cortes@iea-hamburg.de), Überseering 27, 22297 Hamburg, Germany.

*Uncertainty, in the presence of vivid hopes and fears, is painful, but must be endured if we wish to live without the support of comforting fairy tales.*

Bertrand Russell, A History of Western Philosophy (p. 2)

## 1. Introduction

International large-scale assessment of educational achievement (ILSA, hereafter) studies seek to assess education systems in a specific subject domain and in an internationally comparative way. To achieve this, these studies collect data with the objective of generating knowledge about the distribution of student achievement in each of the education systems assessed. It is widely recognized, however, that non-sampling error associated with school non-participation and with non-participation of students at participating schools pose a fundamental threat to the validity of the reported findings.

ILSA studies generally adopt a uniform approach to the problem of non-participation by imposing ignorability assumptions. However, as noted by Brown et al. (2007) in their assessment of four different ILSA studies (p.626): "Non-response bias affecting estimates of central tendency or dispersion for any country is unlikely to be the same across surveys." Furthermore, in their assessment of non-participation patterns in one prominent ILSA study—the Programme for International Student Assessment (PISA)—Durrant and Schnepf (2018) found that the covariates associated with non-participation differed between the school and student levels.

In this article, we examine the problem of non-participation in the context of multi-stage stratified sample designs as typically implemented in ILSA studies. In particular, we bring to the field of ILSA the partial identification, estimation, and reporting strategy advocated by Manski (2003, 2016) that has gained momentum during the last two decades in the field of econometrics. A population parameter of interest is said to be *partially identified*, rather than point identified, when it is known only to lie in a subset of the parameter space even with unlimited sample size.

To account for non-participation, ILSA studies have traditionally focused on employing strategies that obtain point identification of population parameters of interest, such as mean achievement or the share of students demonstrating proficiency. This methodology is appropriate under strong assumptions on the ignorability of non-participation, typically maintained within sampling strata and schools, in which case the population parameter is point identified using data solely from participating schools and students. Partial identification strategies, in contrast, attempt to credibly account for non-sampling error. If the required ignorability assumption is credible, then partial identification analysis leads to the same point identification and estimation results as the conventional approach. If, however, one is concerned that non-participation is not ignorable, then partial identification analysis enables consideration of weaker, but more credible, assumptions that will yield more credible estimates of population parameters.

To illustrate, consider the share of students who demonstrate basic proficiency in math. According to OECD (2023), results from the 2022 PISA indicate that (p. 26) "69% of students are at least basically proficient in mathematics in OECD countries." Abstracting from the multi-stage stratified sample design, suppose that students are selected for the assessment via simple random sampling and that four out of every five sampled students participate, which is reasonably consistent with reported rates. *Without maintaining any assumptions on differences between participants and non-participants*, the overall proficiency rate is partially identified, estimated to fall between 55% and 75%, where the lower bound of the interval corresponds to the case where no non-participants are proficient and the upper bound arises if all non-participants are proficient. Further, *if the incentives to participate are such that it is credible to assert that non-participants tend to be less proficient than are participants*, then the upper bound falls to 69%—that is, the proficiency rate among participants. Finally, *if non-participation is ignorable*, then proficiency rates among participants and non-participants are equal and this estimated upper bound of 69% becomes the point estimate of the mathematics proficiency rate in OECD countries, as stated in the report on 2022 PISA results. This process of assessing the impact of a sequence of assumptions is the essence of a partial identification analysis.

In practice, however, analysis of student non-participation in ILSA studies must address additional subtleties arising from the multi-stage stratified sample design; typically, in the form of a two-stage process. In the first stage, schools are sampled and invited to participate. If a sampled school cannot or chooses not to participate, then no students within this school can participate. In the second stage, within each participating school, students are randomly selected to complete the assessment, but individual participation is voluntary. The identification analysis should credibly account for this multi-stage process.

Regardless of the sampling stage at which it occurs, non-participation reduces the number of students assessed and, in turn, the precision of an estimate based on the data. To focus attention on identification problems, we abstract here from concerns about sampling error associated with non-participation and, in our empirical work, we do not report estimates of sampling precision. Instead, we ask: what can we learn about an education system if the entire population of students were to be selected to participate but some do not take part in the assessment? For discussion of statistical inference of partially identified parameters, see, for example, Canay and Shaikh (2017), Molinari (2020), and Tamer (2010). Vansteelandt et al. (2006) provide a related discussion on the estimation of errors arising from sampling imprecision in regions of ignorance. An ignorance region is a concept used in the sensitivity analysis literature that corresponds to the identification region concept used in the partial identification literature in econometrics. In the course of this paper, we provide some discussion of the overlap between these two methodologies.

We focus attention here on mean achievement within an education system. We recognize the importance of considering not only the central tendency but also the dispersion of educational achievement, as stressed by Brown et al. (2007) who ask whether the data suggest there is a "trade-off between higher mean achievement and lower dispersion." Yet, reports of ILSA results tend to focus on population means and proportions. Further, while partial identification analysis of spread parameters, such as the variance and interquartile range, is feasible, the math is typically more subtle than in the case of the mean on which we focus below and the proficiency rate example above. See, for example, Blundell et al. (2007) and Stoye (2010).

Even with our focus on the population mean and while abstracting from concerns about sampling precision, we trust that this article contributes to the field of ILSA studies and advances research on partial identification to the case of multi-stage stratified sample designs. With respect to ILSA studies, we introduce the framework of partial identification and tailor the approach to account for the sample design that is characteristic of these studies. This framework makes a clear-cut distinction between observable and unobservable information in the estimation. Therefore, it helps determine the boundaries of what can be learned about the probability distribution of student achievement in the assessment from the observable information alone. Importantly, the findings make clear that much can be learned about educational achievement with credible assumptions on the unobservable population.

Using the partial identification framework, we examine the set of assumptions embedded in the standard model applied in ILSA studies to account for non-participation in multiple stages. We show that these assumptions are sufficiently strong to obtain point identification. Yet, the strength of the identifying power may cast doubts on the credibility of the inferences one can reach. We propose and demonstrate the power of maintaining weaker assumptions instead, as represented by the extent to which they shrink the identification region.

These contributions come at a time when ILSA studies—such as the Progress in International Reading Literacy Study (PIRLS) (Mullis et al., 2023), the Trends in International Mathematics and Science Study (TIMSS) (Mullis et al., 2020), and the PISA (OECD, 2023)—are an increasingly important component of the international landscape of educational assessment. This is evident from the number of education systems assessed. For example, 65 education systems took part in the PIRLS assessment in 2021, 72 education systems were assessed by TIMSS in 2023, and about 90 education systems are expected to participate in the PISA in 2025.

There appears to be a consensus among agencies conducting these assessments that sufficiently high student participation rates are critical to producing credible findings. For example, the International Association for the Evaluation of Educational Achievement (IEA) set technical standards that recognize that participation rates reflect "quantitative information (…) to indicate the potential for non-sampling error" (Martin, 1999, pp. 71). To reflect this risk, it is customary to classify populations according to their achieved participation rate and, therefore, implicitly judge the credibility of the reported estimates accordingly.

While participation rates garner notable attention in the applied field of ILSA studies, the strategies to account for them have been extremely limited. At most, to assess the plausibility of the standard model of non-participation, some ILSA studies recommend a so-called non-response bias analysis when participation rates fall below some threshold. The findings on non-response bias may then lead to cautionary annotations being required when reporting assessment results. See, for example, OECD (2024).

The remainder of this article proceeds as follows: Section 2 describes the identification problem arising from student non-participation in the context of ILSA studies when estimating mean achievement. It also discusses several identification strategies, emphasizing the trade-off between ambiguity and credibility. Section 3 demonstrates how to estimate the partially identified parameters within the structure of ILSA studies. In Section 4, we apply this methodology to the 2018 International Computer and Information Literacy Study (ICILS). Specifically, we propose a set of alternative estimation strategies which we believe are reasonable and credible, and we show how these results may be reported by agencies that administer the assessments. Section 5 concludes with suggested extensions to this framework for use in ILSA studies.

## 2. Partial identification of mean achievement

The identification problem arising from student non-participation may be demonstrated as follows. Suppose each student $i$ in an education system is characterized by the duplet $(Y_i, Z_i)$. The variable $Y$ denotes measured student achievement in the assessment and, without loss of generality, achievement is assumed to be measured without error in this paper. The binary variable $Z$ indicates whether a student participates in the assessment. That is, with unlimited sample size, $Z_i$ is observable for each student $i$ in the population, but $Y_i$ is observable only if $Z_i = 1$. This simple structure allows us to use the Law of Iterated Expectations to examine the problem at hand. This statistical law states that the mean of $Y$ can be expressed as the expectation over the conditional expectation of $Y$ given the random variable $Z$. With binary $Z$, we can rewrite $E[Y]$ as,

$$E[Y] = E[Y|Z = 1] \cdot P(Z = 1) + E[Y|Z = 0] \cdot P(Z = 0). \quad 1$$

Clearly, if $P(Z = 0) > 0$ , then $E[Y]$ cannot be point-identified without placing strong assumptions on mean achievement among non-participants $E[Y|Z = 0]$, since this conditional mean is not observable.

However, with the observability of both the probability of participation $P(Z = 1)$ and the mean achievement among participants $E[Y|Z = 1]$, we can deduce that every value within the following identification region $H\{E[Y]\}$ is logically possible for $E[Y]$:

$$H\{E[Y]\} = E[Y|Z = 1] \cdot P(Z = 1) + \phi \cdot P(Z = 0), \;\; \phi \in \mathrm{T}_{E[Y]}, \quad 2$$

where $\mathrm{T}_{E[Y]}$ is the set of all possible values for the conditional mean of $Y$ within the subpopulation of students with $Z = 0$. This identification region is informative about the population mean—that is, the population mean is partially identified—to the extent that some restrictions may be placed on logically possible values for the unobservable $E[Y|Z = 0]$, such as in the case where the distribution of $Y$ is known to have bounded support.

In the remainder of this section, we present a partial identification analysis considering how different restrictions on $E[Y|Z = 0]$ impact the conclusions that can be reached about $E[Y]$. It is worth noting that some of the results presented here can also be derived from the class of models introduced by Rotnitzky et al. (1998), Robins et al. (2000), and Rotnitzky et al. (2001). Under these models, $E[Y]$ can be nonparametrically point-identified by specifying a sensitivity function that assumes knowledge about $P(Z = 1|Y)$. Scharfstein et al. (2004) provide a notable example about how identification analysis and sensitivity analysis can be juxtaposed in longitudinal studies with attrition. In the context of this paper, sensitivity analysis requires strong assumptions about the (unknown) sensitivity function determining how the probability of participation in the assessment varies with student proficiency in the assessment. In the discussion below, we focus our attention on partial identification analysis. Where appropriate, we provide an interpretation of the nonparametric bounds derived using the partial identification framework through the lens of the sensitivity analysis framework.

### 2.1 Minimal restrictions on $E[Y|Z = 0]$

As advocated by Manski (1995), a natural starting point in partial identification analysis is to determine the limits of what can be learned through observation alone or while imposing minimal restrictions on the unobservables, such as regularity conditions or knowledge that the unknown distribution has bounded support. A useful starting point in our application is to suppose $E[Y|Z = 0]$ is known to fall within some range such that $Y^L \leq E[Y|Z = 0] \leq Y^U$, where the values $Y^L$ and $Y^U$ are known lower and upper limits, such as the limits of an assessment scale.

This restriction generates sharp bounds on $H\{E[Y]\}$. That is, with $Y^L \leq E[Y|Z = 0] \leq Y^U$, $E[Y]$ is known to satisfy the following expression,

$$E[Y|Z = 1] \cdot P(Z = 1) + Y^L \cdot P(Z = 0) \leq E[Y] \leq E[Y|Z = 1] \cdot P(Z = 1) + Y^U \cdot P(Z = 0) \quad 3$$

To gauge the informativeness of these bounds, we note that this identification region has width $P(Z = 0) \cdot \left(Y^U - Y^L\right)$.

*Distributions with bounded support*

In some cases, there may be natural lower or upper limits on the support of $Y$. For instance, if $Y$ were a binary variable indicating whether student achievement exceeds a threshold value—e.g., the value that is indicative of basic proficiency in mathematics—then it would follow that $Y^L = 0$ and $Y^U = 1$ and the identification region has width $P(Z = 0)$, as in the math proficiency example in the Introduction where $P(Z = 0) = 0.20$. Similarly, if $Y$ were the number of correct answers given by a student, then it would follow that $Y^L = 0$ and $Y^U$ is the number of test items.

*Distributions with unbounded support*

However, in the context of achievement scores in ILSA, the scale on which scores are reported has no natural bounds. In the first cycle of a study, it is standard practice for scale scores to have a specific international average and standard deviation. To be able to measure trends in subsequent cycles, scores are anchored to the scale from the first cycle of assessments.

Credible values of $Y^L$and $Y^U$ may be context dependent and specific to the education system that is being assessed. In the absence of obvious a priori values for $Y^L$ and $Y^U$, it may be credible to use the observable distribution of $Y$ (i.e. among those with $Z = 1$) to restrict $E[Y|Z = 0]$ via a *bounded-variation assumption*. For example, one may assert that mean achievement among non-participating students does not fall within the tails of the achievement distribution among all participating students, as we consider here with Assumption 1.

**Assumption 1 (A1)**: Mean achievement of the non-participating students in an education system lies within an interval bounded by the $\alpha_L^{th}$ and $\alpha_U^{th}$ percentiles of the achievement distribution among the participating students, where $0 < \alpha_L \leq \alpha_U < 1.0$.

Assumption A1 places no substantive restriction on the values of $\alpha_L$ and $\alpha_U$. Its credibility, however, depends crucially on the selected values. For instance, asserting that $\alpha_L = \alpha_U = 0.50$ would generate point identification of $E[Y|Z = 0]$, in which case the strength of the assumption is comparable to that of assuming that non-participation is ignorable. In fact, the assumptions would be identical if the distribution of $Y$ among participating students were known to be symmetric. In contrast, values for $(\alpha_L, \alpha_U)$ of $(0.05, 0.95)$ or, perhaps, $(0.10, 0.90)$ would constitute much weaker assumptions, typically allowing mean achievement scores to vary quite substantially between participants and non-participants. The extent of the potential deviation depends on both the choice of $(\alpha_L, \alpha_U)$ and the spread of the distribution of achievement scores among participants.

Assumption A1 may be written as $Q_{\alpha_L}(Y|Z = 1) \leq E[Y|Z = 0] \leq Q_{\alpha_U}(Y|Z = 1)$, where $Q_{\alpha}(Y|Z = 1)$ denotes the $\alpha$-quantile, or $\alpha^{th}$ percentile, of the probability distribution of $Y$ among participating students. Under assumption A1, $H\{E[y]\}$ has sharp bounds that can be expressed as follows:

$$
\begin{gathered}
E[Y|Z = 1] \cdot P(Z = 1) + Q_{\alpha_L}(Y|Z = 1) \cdot P(Z = 0) \\
\leq E[Y] \leq \\
E[Y|Z = 1] \cdot P(Z = 1) + Q_{\alpha_U}(Y|Z = 1) \cdot P(Z = 0)
\end{gathered} \qquad 4
$$

Assumption A1 therefore allows us to characterize $H\{E[y]\}$ with information that can be established through observation.

Importantly, the width of $H\{E[y]\}$ under assumption A1 is given by $P(Z=0)\cdot\left[\left(Q_{\alpha_U}(Y|Z=1)-Q_{\alpha_L}(Y|Z=1)\right)\right]$, which captures the ambiguity about $E[y]$ under this assumption. Choosing a smaller value for $\alpha_L$ or larger value for $\alpha_U$ may lead to more credible, but also more ambiguous, findings on $E[y]$.

In the context of ILSA, the values chosen for $(\alpha_L, \alpha_U)$ could vary from context to context. For example, OECD (2022) reports that an average OECD country has a mean achievement score in the PISA math assessment of 472 points with a standard deviation of 90. If these scores are normally distributed, then choosing values of $(0.05, 0.95)$ would imply that mean achievement among non-participants $E[Y|Z=0]$ lies somewhere within the interval $[324, 620]$; that is, up to 148 points below or above the mean achievement score among participants. This assumption appears to be credible but perhaps far weaker than necessary given evidence found in non-response bias analyses of PISA scores. For example, OECD (2022, p. 17-20) indicates that, in most countries, the non-participating population would score, on average, about ten points less than the participating population. These estimates of potential bias are compatible with findings reported for PISA scores in England by Micklewright et al. (2012, pp. 932), who report "upward bias in the estimates of the mean from the achieved sample of respondents of about 7—9 points in 2003" and "between 4 and 15 points" in 2000.

*Partial identification analysis in stratified designs with minimal restrictions*

An important design feature of ILSA studies is school stratification in the assessment. School stratification divides the universe of schools in an education system into a set of $W$ groups or strata, where each school is assigned to one and only one school stratum according to some observable characteristics. Provided that each student in a population is enrolled in only one school, school stratification also divides the universe of students into a set of $W$ strata. This partition seeks to group—with respect to the assessment outcomes—a potentially heterogeneous education system into subgroups, each of which being more homogenous internally.

For ILSA studies, school stratification has two main aims. The first aim is to increase efficiency in estimation by independently sampling schools to participate in the assessment within homogenous strata. The second aim is of direct relevance to identification. That is, school stratification serves as a building block in the standard model of non-participation, where student non-participation arising from school-level non-participation is assumed to be ignorable within each stratum $w$.

Recognizing the potential information content of observed school stratum with respect to student achievement, adjusting assumption A1 to be stratum-specific may, perhaps, lead to both more credible and less ambiguous findings. To incorporate this design feature into our identification analysis, we can rewrite $E[Y]$ as (see, for example, Chapter 5 and Chapter 5A in Cochran (1977)),

$$E[Y]=\sum_{w=1}^{W}P(W=w)\cdot E[Y|W=w] \qquad 5$$

where $P(W=w)$ is the event probability that a student belongs to a school in stratum $w$. Expression 5 gives sufficient structure for the following assumption,

**Assumption 1.1 (A1.1)**: Mean achievement of the non-participating students in each stratum $w$ lies within an interval bounded by the $\alpha_L^{th}$ and $\alpha_U^{th}$ percentiles of the achievement distribution among the participating students in that stratum, where $0 < \alpha_L \leq \alpha_U < 1.0$.

It follows that under assumption A1.1, $H\{E[y]\}$ is fully characterized by the values satisfying the following expression,

$$\sum_{w=1}^{W} P(W=w)\Big(E[Y|Z=1,W=w]\cdot P(Z=1|W=w) + Q_{\alpha_L}(Y|Z=1,W=w)\cdot P(Z=0|W=w)\Big)$$
$$\leq E[Y] \leq \qquad 6$$
$$\sum_{w=1}^{W} P(W=w)\Big(E[Y|Z=1,W=w]\cdot P(Z=1|W=w) + Q_{\alpha_U}(Y|Z=1,W=w)\cdot P(Z=0|W=w)\Big)$$

Invoking assumption A1.1 rather than A1 may lead to both more credible inference and less ambiguity if the distribution of achievement is known to vary systematically across strata. In fact, Durrant and Schnepf (2018) suggest that school observable characteristics, such as proxies for socio-economic status, are informative to predict school participation in the assessment and therefore incorporating them in the analysis may lead to more credible conclusions.

## 2.2 Standard assumptions in ILSA

In the remainder of this section, we demonstrate how different assumptions about the unobservable $E[Y|Z=0]$ allow us to learn about $E[Y]$ beyond the limits revealed by Expression 6. We begin by examining the standard model applied in ILSA studies to account for non-participation in the assessment, where ignorability assumptions within some adjustment cells are maintained. These studies assert that non-participants are missing at random (MAR) conditional on observed covariates (Rubin, 1976). In ILSA studies, the conditioning information is typically restricted to information that is available during sampling and data collection. That is, the assumption that school participation is at random is typically conditioned on school membership to stratum $w$, while the assumption that student participation is at random is typically conditioned on school membership (see, for example OECD (2024) and Tieck (2020b)).

To more fully reflect the structure of an ILSA design and therefore the stages at which non-participation can occur in these assessments, it will be useful to expand the notation introduced in Section 2.1. Let each student $i$ in a population be characterized by the triplet $(Y_i, Z_{1i}, Z_{2i})$, where $Z_{1i}$ denotes whether student $i$ attends a school that would participate in the study if sampled, and $Z_{2i}$ indicates whether student $i$ would participate in the assessment if sampled from within their participating school. In this context, a student can only participate in the assessment if their school participates. Thus, the binary indicator $Z_i = 1$ if and only if both $Z_{1i} = 1$ and $Z_{2i} = 1$; otherwise $Z_i = 0$. To incorporate this structure, we write the stratum-level mean of $Y$ by using the Law of Iterated Expectations,

$$\begin{aligned} E[Y|W=w] &= E[Y|Z_1=1,Z_2=1,W=w]\cdot P(Z_1=1,Z_2=1,W=w) \\ &+E[Y|Z_1=1,Z_2=0,W=w]\cdot P(Z_1=1,Z_2=0,W=w) \\ &+E[Y|Z_1=0,W=w]\cdot P(Z_1=0,W=w). \end{aligned} \qquad 7$$

Thus, for each stratum $w$, we partition the population into three mutually exclusive and exhaustive subgroups: $(Z_1=1,Z_2=1)$ participating students in participating schools, $(Z_1=1,Z_2=0)$ non-participating students in participating schools, and $(Z_1 = 0)$ students in non-participating schools. We see in Expression 7 that each of the three event probabilities are observable, but only the first of the three conditional means is observable—that is, $E[Y|Z_1 = 1, Z_2 = 1, W = w]$ the mean achievement of participating students in participating schools in stratum $w$.

Now consider the second conditional mean $E[Y|Z_1 = 1, Z_2 = 0, W = w]$, mean achievement of non-participating students in a participating school in stratum $w$. To account for student non-participation within participating schools, it is common practice in ILSA to use school units as so-called adjustment cells. Correspondingly, the following assumption is typically maintained:

**Assumption 2 (A2)**: Student achievement is statistically independent of student participation within school $j$, for all schools $j$.

This assumption means that, within each participating school, student non-participation is at random and leads to the result that $E_j[Y|Z_1 = 1, Z_2 = 1, W = w] = E_j[Y|Z_1 = 1, Z_2 = 0, W = w]$ for all schools $j$. Under assumption A2, Expression 7 reduces to:

$$\begin{aligned} E[Y|W = w] = E[Y|Z_1 = 1, Z_2 = 1, W = w] \cdot P(Z_1 = 1, W = w) \\ +E[Y|Z_1 = 0, W = w] \cdot P(Z_1 = 0, W = w) \end{aligned} \quad 8$$

Note that under assumption A2, school-level non-participation is the only remaining source of non-participation that prevents point-identification of $E[Y|W = w]$.

Recognizing that one may accept as credible that student non-participation within participating schools is ignorable, as implied by A2, but that ignorability of school-level non-participation is not credible, we propose the following modification to assumption A1.1 to specify credible bounds on student achievement in non-participating schools:

**Assumption 1.2 (A1.2)**: Mean achievement of students in non-participating schools, in each stratum $w$, lies within an interval bounded by the $\alpha_L^{th}$ and the $\alpha_U^{th}$ percentiles of the achievement distribution among students in participating schools in that stratum, where $0 < \alpha_L \leq \alpha_U < 1.0$.

It follows that, under assumptions A1.2 and A2, $H\{E[Y]\}$ is fully characterized by the values satisfying the following expression,

$$\begin{gathered} \sum_{w=1}^{W} P(W = w) \cdot \begin{pmatrix} E[Y|Z_1 = 1, Z_2 = 1, W = w] \cdot P(Z_1 = 1, W = w) \\ +Q_{\alpha_L}(Y|Z_1 = 1, Z_2 = 1, W = w) \cdot P(Z_1 = 0, W = w) \end{pmatrix} \\ \leq E[Y] \leq \\ \sum_{w=1}^{W} P(W = w) \cdot \begin{pmatrix} E[Y|Z_1 = 1, Z_2 = 1, W = w] \cdot P(Z_1 = 1, W = w) \\ +Q_{\alpha_U}(Y|Z_1 = 1, Z_2 = 1, W = w) \cdot P(Z_1 = 0, W = w) \end{pmatrix} \end{gathered} \quad 9$$

To account for the fact that some schools do not take part in the assessment, rather than a potentially weak assumption such as A1.2., ILSA studies invoke ignorability assumptions on school non-participation, whereby the information used for stratifying the school sample is leveraged to assert that schools are missing is at random within each stratum $w$, as follows:

**Assumption 3 (A3)**: Student achievement is statistically independent of school participation within sampling stratum $w$, for all strata $w$.

Assumption 3 leads to the result that $E[Y|Z_1 = 0, W = w] = E[Y|Z_1 = 1, W = w]$ for all $w$. Under this *additional* assumption, stratum-specific mean achievement reduces to $E[Y|W = w] = E[Y|Z_1 = 1, Z_2 = 1,\ W = w]$ for all strata $w$. We therefore conclude that, under assumptions A2 and A3, $E[Y]$ is point identified:

$$E[Y] = \sum_{w=1}^{W} P(W = w)E[Y|Z_1 = 1, Z_2 = 1, W = w] \qquad 10$$

That is, $H\{E[Y]\}$ is fully characterized by the singleton in Expression 10, illustrating the power of assumptions A2 and A3. Note further that if $Q_{\alpha_L}(Y|Z = 1, W = w) \leq E[Y|Z = 1, W = w] \leq Q_{\alpha_U}(Y|Z = 1, W = w)$ for all strata $w$—which should be true for any credible choice of $(\alpha_L, \alpha_U)$ in an ILSA study in which A2 and A3 are plausible—then $E[Y]$ under assumptions A2 and A3 will fall within the identification region of $E[Y]$ under assumption A1.1, characterized above in Expression 6. A similar result holds for the identification region in Expression 9 under assumptions A1.2 and A2.

Interestingly, in a sensitivity analysis, the same result could be achieved by maintaining the assumptions that (i) within each school $j$ the sensitivity function capturing student participation in the assessment is a constant and (ii) within each stratum $w$ the sensitivity function capturing school participation is a constant. Such assumptions imply that, within each participating school $j$, $P_j(Z_{2i}|Y_i,) = P_j(Z_{2i})$ and that, within each stratum $w$, $P_w(Z_{1i}|Y_i) = P_w(Z_{1i})$, respectively. In other words, missingness is at random within each adjustment cell.

*Discussion of ignorability assumptions*

As noted above, ILSA studies typically utilize a limited set of school and student information on which to condition the ignorability of missingness—in particular, school stratum for school-level non-participation and students' school membership for within-school non-participation (see, for example, ICILS 2018 Technical Report (Tieck, 2020b, pp.80-81) and PISA 2022 Technical Report (OECD, 2024, pp. 191-192)). The discussion above could be easily modified to allow for additional covariates $X$. In fact, some may believe that adding more covariates will make the ignorability assumption more plausible. As Little (2021) wrote, referring to the highly influential work of Donald Rubin (p. 102): "Rubin himself has argued that with a sufficiently rich set of observed data, MAR is often justified." As also noted by Little in the preceding sentence, however: "The problem is that we usually cannot tell from the observed data whether or not MAR applies."

Further, Manski (2007) argued (pp. 65-66):

"There is no general foundation to the assertion that missingness at random becomes a better assumption as one conditions on more covariates. The assertion may be well grounded in some settings, but it is not self-evident. Indeed, outcomes may be missing at random conditional on $x$ but not conditional on the longer covariate vector $(x, w)$."

*Discussion of replacement schools*

Finally, an important component of the model commonly applied in ILSA studies to account for non-participation is the use of replacement schools, which corresponds to a data imputation technique that resembles the nearest-neighbor hot-deck method. This mechanism is put into place when a sampled, non-participating school is "replaced" by a school that was selected specifically for the purpose of replacing the non-participating school. That is, a "recipient school" is replaced by a "replacement school". The data collected in a replacement school are then used to estimate population parameters as if the data came from the recipient school. Martin et al. (1999) and Tieck (2020) provide a detailed discussion about how replacement schools are matched to recipient schools.

Incorporating this component of the non-participation model into an identification analysis requires some subtlety because identification analyses proceed as if all students in an education system are sampled. Thus, no non-sampled schools would be available to replace sampled, non-participating schools. One may instead consider hypothetically drawing replacement schools from the pool of sampled and participating schools, as in hot-deck imputation. However, this imputation serves no obvious purpose for identification. That is, replacement schools only provide information about students in the education system that would participate whenever sampled, but they are uninformative about the portion of students in schools that would not participate when sampled. One might then ask, what purpose is served by using replacement schools? The purpose is to increase sample size and sampling precision when engaging in an exercise of statistical inference. From the perspective of identification, assumptions A2 and A3 are sufficient to point identify $E[Y]$.

### 2.3 Monotone non-participation

Finally, we present an alternative identification strategy that incorporates a plausible assumption of monotone non-participation. Assumption 4 below is compatible with a model of participation where there is some achievement-related incentive for (non)participation in the assessment. For example, it may be credible to assert that schools with lower achieving students are less likely to agree to participate when sampled and/or that lower achieving students are less likely to participate in an assessment than are their classmates.

**Assumption 4 (A4)**: Mean achievement in the participating subpopulation is weakly greater than mean achievement in the non-participating subpopulation within sampling stratum $w$, for all strata $w \in W$.

This assumption implies $E[Y|Z=0, W=w] \leq E[Y|Z=1, W=w]$ for all $w$. The credibility of assumption A4 is supported by evidence collected in PISA non-response bias analyses described above (OECD, 2022; Micklewright et al., 2012).

Suppose further that, $E[Y|Z=1, W=w] \leq Q_{\alpha_U}(Y|Z=1, W=w)$, for all strata $w$. Then, putting assumptions A1.1 and A4 together, the identification region of interest is expressed by the set of values that satisfy the following,

$$\sum_{w=1}^{W} P(W=w) \cdot \big(E[Y|Z=1, W=w] \cdot P(Z=1|W=w) + Q_{\alpha}(Y|Z=1, W=w) \cdot P(Z=0|W=w)\big)$$
$$\leq E[Y] \leq$$
$$\sum_{w=1}^{W} P(W=w) \cdot E[Y|Z=1, W=w] \qquad 11$$

It is worth noting that the lower bound in Expression 11 coincides with the lower bound of Expression 6, while the upper bound of Expression 11 is the strata-weighted average of the means of scores among participating students.

## 3. Interval estimation of a partially identified mean

In the previous section, we demonstrated how to derive different identification regions of $E[Y]$ by invoking assumptions on the achievement distribution among non-participating students. In each instance, we identified sharp bounds that determine the set of logically possible values that $E[Y]$ can take under the maintained set of assumptions. The identified bounds are functions of observable population parameters. We now discuss the sample counterparts, or "analogs," of the identification regions derived in the previous section. We may then apply the "analogy principle" to use these sample statistics to estimate the analogous population parameters as described by Goldberger (1991, pp. 117): "The most natural rule for selecting an estimator is the *analogy principle*. A population parameter is a feature of the population. To estimate it, use the corresponding features of the sample.".

Randomization in the sample design allows us to generate partial knowledge about $E[Y]$ from a subset of the student population. The design typically implemented in ILSA studies follows a two-step process that can be summarized as follows. In the first stage, within each stratum, schools are sampled with probabilities proportional to their relative size in their stratum. This implies that a measure of size $M_j$ is known for each school $j$ in the population and that larger schools have a higher probability of selection. As a norm, the measure of school size is the enrollment of students belonging to the target population of the study (Martin et al., 1999). In the second stage, within participating schools, students or intact classes are selected to participate in the assessment with equal probabilities. Such a sampling plan leads to probabilities of inclusion that vary across sampled students. Therefore, the probability that student $i$ in school $j$ is sampled for the assessment is given by the joint probability $\pi_{ij} = \pi_j * \pi_{i|j}$ where $\pi_j$ denotes the probability that school $j$ is sampled and $\pi_{i|j}$ denotes the probability that student $i$ is sampled from within their sampled school $j$.

In what follows, we discuss estimation of various population parameters characterizing the non-parametric bounds derived in the previous section. For estimation, we use inverse probability weighting (IPW), for which different known or assumed probabilities are used as propensity scores. We restrict attention to IPW methods, as these are typically used for reporting in ILSA studies. We note, however, that IPW methods are generally inefficient due to the substantial variation in the propensity scores. For example, substantial variation in $\pi_j$ and $\pi_{i|j}$ across participating students can lead to substantial sampling errors. Graham et al. (2012) provides an instructive discussion about the properties of IPW estimators.

### 3.1 Minimal restrictions on $E[Y|Z = 0]$

Estimation of the bounds that characterize $H\{E[Y]\}$ under maintained assumption A1 (i.e., Expression 4) requires estimation of several features of the distribution of achievement among participating students—$E[Y|Z = 1]$, $Q_{\alpha_L}(Y|Z = 1)$, and $Q_{\alpha_U}(Y|Z = 1)$—as well as the event probability of participation. To estimate $P(Z = 1)$ under the implemented sampling plan, it is useful to recall that we can characterize each student $i$ in our population of interest by the triplet $(Y_i, Z_{1ij}, Z_{2i})$; where $Z_{1ij}$ denotes whether student $i$ attends a school $j$ that would participate in the study if sampled, and $Z_{2i}$ indicates whether student $i$ would participate in the assessment if sampled from within their participating school. In this context, a student can only participate in an assessment if their school participates. Thus, the binary indicator $Z = 1$ if and only if both $Z_1 = 1$ and $Z_2 = 1$; $Z = 0$ otherwise (see Section 2.2). It follows that if $P(Z_1 = 1) > 0$ then $P(Z = 1) = P(Z_1 = 1, Z_2 = 1) = P(Z_1 = 1) \cdot P(Z_2 = 1 | Z_1 = 1)$.

Given the two-stage nature of the implemented sampling plan, decomposing $P(Z = 1)$ into these two components is useful for representing the event probability that a student participates in ILSA studies. The first stage of the sampling plan generates knowledge about $P(Z_1 = 1)$, while the second stage generates knowledge about $P(Z_2 = 1 | Z_1 = 1)$.

To estimate $P(Z_1 = 1)$ we note that, from a frequentist perspective, the probability of an event is nothing other than the long-run rate at which the event occurs. Therefore, we propose using its sample analog $\hat{p}_1$,

$$\hat{p}_1 = \frac{\sum_{j=1}^{J} \pi_j^{-1} M_j z_{1j}}{\sum_{j=1}^{J} \pi_j^{-1} M_j}. \qquad 12$$

where $j = 1, \ldots, J$ indexes sampled schools. That is, the following information is available for each sampled school $j$: $z_{1j}$ which indicates its participation status in the study; $M_j$, which denotes the number of students enrolled in it; and its sampling weight $\pi_j^{-1}$, which is given by the inverse of its inclusion probability. Note that the denominator in Expression 12 estimates the total number of students enrolled in the population, while the numerator estimates the total number of students enrolled in schools that would participate in the study if sampled. Therefore, $\hat{p}_1$ estimates the proportion of students enrolled in schools that would participate in the study if sampled.

Similarly, to estimate $P(Z_2 = 1 | Z_1 = 1)$ we propose using its sample analog $\hat{p}_2$,

$$\hat{p}_2 = \frac{\sum_{j=1}^{J} \sum_{i=1}^{N_j} \pi_{ij}^{-1} z_{1j} z_{2ij}}{\sum_{j=1}^{J} \sum_{i=1}^{N_j} \pi_{ij}^{-1} z_{1j}} \qquad 13$$

here $i = 1, \ldots, N_j$ indexes sampled students within sampled school $j$. That is, the following information is available for each sampled student $i$ in participating school $j$: $z_{2ij}$, which indicates the student's participation status in the assessment, as well as the student's sampling weight $\pi_{ij}^{-1}$, which is given by the inverse of the inclusion probability. These values are not observed for student $i$ in *non-participating* school $j$, but that is not problematic because we observe $z_{1j} = 0$ and, therefore, the summand in the numerator and the summand in the denominator are each known to equal zero for each non-participating school $j$. Note that the denominator in Expression 13 estimates the total number of students enrolled in schools that would participate if sampled, while the numerator estimates the total number of students that would participate in the assessment if sampled across the schools that would participate if sampled. Therefore, $\hat{p}_2$ estimates the proportion of students that would participate in the assessment if sampled, across schools that would participate if sampled. It is worth noting that the denominator in Expression 13 and the numerator in Expression 12 each estimate the same parameter but with different data. Both estimates coincide if $N_j = M_j$, which is usually not the case. This is because observation of $N_j$ occurs during data collection in participating schools, which happens at least one year later than the observed enrollment used for the construction of the school frame ($M_j$).

Thus, to estimate $P(Z = 1)$ we propose using its sample analog $\hat{p}$,

$$\hat{p} = \hat{p}_1 * \hat{p}_2 \qquad 14$$

Similarly, to estimate $E[Y|Z = 1]$, we propose to use the sample analog $\hat{\mu}$, a weighted average of the assessment outcomes $y_{ij}$ observed in the sample,

$$\hat{\mu} = \frac{\sum_{j=1}^{J}\sum_{i=1}^{N_j}\pi_{ij}^{-1}y_{ij}z_{1j}z_{2ij}}{\sum_{j=1}^{J}\sum_{i=1}^{N_j}\pi_{ij}^{-1}z_{1j}z_{2i}}. \qquad 15$$

where $y_{ij}$ denotes achievement of sampled student $i$ in school $j$. Note that $y_{ij}$ is observed if and only if $z_{2ij} = 1$, and both $\pi_{ij}^{-1}$ and $z_{2ij}$ are observed if and only if $z_{1j} = 1$. As above, in cases where any of these values are not observed, $z_{1j} = 0$ and, therefore, the summand in the numerator and the summand in the denominator are each known to equal zero as well.

Analog estimation of percentiles of the distribution of $Y$ among students characterized by $Z = 1$proceeds accordingly, using percentiles of the distribution of achievement outcomes observed in the sample. Let $\hat{q}_\alpha$ denote the analog estimate of the $\alpha^{th}$ percentile $Q_\alpha(Y|Z = 1)$. Then, the sample counterpart of the bounds in Expression 4 is then given by,

$$\left[\hat{\mu}\cdot\hat{p} + \hat{q}_{\alpha_L}\cdot(1-\hat{p}), \hat{\mu}\cdot\hat{p} + \hat{q}_{\alpha_U}\cdot(1-\hat{p})\,\right]. \qquad 16$$

Note that all components of the bounds are estimated using only originally sampled schools; that is, we do not use the so-called replacement schools (see discussion at the end of Section 2.2).

As discussed in Section 2.1, an alternative strategy to make estimation more credible is to incorporate school stratification in the analysis and thus replace assumption A1 with A1.1. If so, the sample counterpart of the bounds in Expression 6 is given by,

$$\left[\sum_{w=1}^{W} P(w)\left(\hat{\mu}_w\hat{p}_w + \hat{q}_{\alpha_L,w}\cdot(1-\hat{p}_w)\right), \sum_{w=1}^{W} P(w)\left(\hat{\mu}_w\hat{p}_w + \hat{q}_{\alpha_U,w}\cdot(1-\hat{p}_w)\right)\right] \qquad 17$$

where $\hat{\mu}_w$, $\hat{q}_{\alpha,w}$, and $\hat{p}_w$ are stratum-specific estimates analogous to those in Expression 16. $P(w)$ is the known proportion of students in the population attending schools in stratum $w$, as reported in the school frame used for sampling. Thus, the bounds in Expression 17 are an average of the stratum-specific bounds, weighted by a factor proportional to the total enrollment in each stratum.

### 3.2 Standard assumptions in ILSA

Now we turn our attention to estimation under the standard model to account for non-participation in ILSA studies. We begin by estimating the bounds of $H\{E[Y]\}$ under assumptions A1.2 and A2, where student non-participation $Z_2$ is ignorable but school non-participation $Z_1$ may not be. The sample counterpart of the bounds in Expression 9 is given by,

$$\left[\sum_{w=1}^{W} P(w)\left(\hat{\mu}_w\hat{p}_w + \hat{q}_{\alpha_L,w}\cdot(1-\hat{p}_{1w})\right), \sum_{w=1}^{W} P(w)\left(\hat{\mu}_w\hat{p}_w + \hat{q}_{\alpha_U,w}\cdot(1-\hat{p}_{1w})\right)\right] \qquad 18$$

where $P(w)$ is as defined in Expression 17. To incorporate assumption A2 into the IPW estimation of $\hat{\mu}_w$, $\hat{q}_{\alpha_L,w}$, and $\hat{q}_{\alpha_U,w}$, we use estimation weights of the form $\omega_{ij} = \left(\pi_{ij} * \pi_j^f\right)^{-1}$. That is, the probability of inclusion $\pi_{ij}$ is adjusted by a school-level factor $\pi_j^f = \frac{\sum_{i=1}^{N_j} z_{2i}}{\sum_{i=1}^{N_j} 1}$, where the numerator reflects the number of participating students in participating school $j$ and the denominator reflects the number of sampled students within the same school. This adjustment factor is an artifact of assumption A2, and it follows from it that the relative contribution of participating students in school $j$ to the overall estimate equals the relative contribution of sampled students in school $j$ to the overall estimate. Finally, since under assumption A2 school-level non-participation is the only source of non-participation that remains in the expression, the length of each of the estimated stratum-specific bounds is proportional to $\hat{p}_1$ rather than $\hat{p}$.

Finally, under maintained assumptions A2 and A3, $H\{E[Y]\}$ is point-identified and its sample counterpart is given by the mean achievement across all participating students, weighted by a factor $\omega^*{}_{ij} = \left(\pi_{ij}\pi_w^f\pi_j^f\right)^{-1}$. That is,

$$\hat{\mu} = \frac{\sum_{j=1}^{J}\sum_{i=1}^{N_j}\omega_{ij}^* y_{ij} z_{1j} z_{2ij}}{\sum_{j=1}^{J}\sum_{i=1}^{N_j}\omega_{ij}^* z_{1j} z_{2ij}} \qquad 19$$

Similar to the above, the inclusion probability $\pi_{ij}^{-1}$ is adjusted by a stratum-level factor $\pi_w^f = \frac{\sum_{j\in w} z_{ij}}{\sum_{j\in w} 1}$, where the numerator reflects the number of participating schools in stratum $w$ and the denominator reflects the number of sampled schools within the same stratum. This adjustment factor is an artifact of assumption A3, and it follows from it that the relative contribution of participating schools in stratum $w$ to the overall estimate equals the relative contribution of sampled schools in stratum $w$ to the overall estimate. It is worth noting that, in the field of ILSA, $\pi_h^f$ and $\pi_h^f$ are known as the school-level and the student-level non-response adjustment factors, respectively.

As noted at the end of Section 2.2, assumptions A2 and A3 are jointly sufficient to point-identify and hence point-estimate $E[Y]$. Yet, ILSA studies typically use replacement schools to increase sample size and sampling precision when engaging in an exercise of statistical inference. One could, alternatively, maintain assumption A2 and A3 after the inclusion of replacement schools. Such a strategy has been chosen by ILSA studies to estimate and report $E[Y]$, where estimation takes place as in the preceding paragraph except that data from some non-participating schools are imputed with data from replacement schools that were not selected in the original sample. The addition of replacement schools leads to a different school-level adjustment factor ($\pi_w^f$) and, hence, weight in the estimation, reflecting the increase in the number of participating schools.

### 3.3 Monotone non-participation

Now we consider estimation of the logically possible values that $E[Y]$ can take under monotone non-participation as presented in Section 2.3. Here, we rule out the extreme case in which $Q_{\alpha_U}(Y|Z=1) < E[Y|Z=1]$. The sample counterpart of the bounds in the identification region under assumptions A1.1 and A4 (Expression 11), is given by,

$$\left[\sum_{w=1}^{W} P(w)\left(\hat{\mu}_w \cdot \hat{p}_w + \hat{q}_{\alpha_L,w} \cdot (1-\hat{p}_w)\right), \sum_{w=1}^{W} P(w)\hat{\mu}_w\right] \qquad 20$$

where components are defined analogously to those in Expression 17.

## 4. Empirical Findings from the 2018 ICILS

In this section, we use the methods developed in the preceding sections to conduct an international comparison of achievement in the 2018 International Computer and Information Literacy Study. This study, coordinated by the IEA, assesses outcomes on the computer and information domain among eighth graders in different education systems. Within each system, ICILS seeks to collect data from a representative sample of students, using a sample design similar to the one described above. For each student participating in the study, ICILS reports an achievement measure of computer and information literacy (CIL), which is a latent construct defined as the "ability to use computers to investigate, create and communicate in order to participate effectively at home, at school, in the workplace and in society'' (Fraillon et al., 2019, pp. 53). We use this measure as the outcome of interest $Y$.

The assessment framework behind the CIL latent construct developed for ICILS is extensive. Overall, four computer-based modules, each with between 15 and 22 tasks, were developed to evaluate literacy in the computer and information domain. To avoid overburdening participating students, the assessment modules are paired, and students are exposed to only one of the 12 possible ordered-paired modules. Therefore, each student responds to only a subset of the overall task pool needed to evaluate CIL. Accordingly, ICILS is designed to assess groups of students rather than individuals. After data are collected and scored, item response theory models are used to estimate the proficiency distribution of the CIL construct for each student (see, for example, Ockwell et al. (2020) for details). In particular, for each student, the ICILS data file includes five plausible values for that student's measured CIL, where each plausible value consists of a random draw from their estimated proficiency distribution. As is customary in ILSA studies, we use the reported sets of five plausible values as a measure of CIL. To do so, we estimate parameters of interest five times, each with one plausible value, and report the average over the five estimates. Von Davier et al. (2009) provides a general discussion of the underlying methodology of this type of assessment design and how to incorporate it into the estimation of population parameters. In the remainder of this section, we follow this standard practice in the estimation of population parameters but do not address this issue in our discussion, as it is a measurement concern that is beyond the scope of our focus here on student non-participation.

We also reiterate that in this section, as in the theoretical discussion above, we abstract from the problem of statistical inference associated with finite sample size. Therefore, the reported estimates do not include any measure of sampling precision, as typically represented by confidence intervals. Thus, here we treat the sample as if it were the population and ask the following question: what can we possibly learn about student mean achievement in an education system from the observed data if all students were selected to participate but some do not take part in the assessment?

### 4.1 Reporting strategy under assumption A1.1

As noted above, there is evidence in many contexts that, on average, the non-participating population would have scored similarly or lower than the participating population if they had participated in the assessment (OECD, 2024; Micklewright, 2012). Therefore, it is desirable to maintain assumptions that reflect this possibility when reporting inferences about $E[Y]$. Interval estimates that conform to this argument can be constructed using assumption A1.1. The question arises, however, as to what percentiles of the observable distribution of $Y$ could be used to derive credible and reasonable lower and upper bounds for $E[Y]$.

First, the evidence of PISA cited above suggests that, in many contexts, if non-participants participated in the assessment, then they would score, on average, about seven to ten points less than participants if they were to participate in the assessment. However, as this evidence comes from sources other than ICILS, which measured a different set of skills, a more conservative approach is appropriate. We believe it is reasonable and credible to assume that the mean achievement of non-participants is weakly larger than the 25th percentile of the achievement distribution among participants, i.e., $Q_{25}(Y|Z=1) \leq E[Y|Z=0]$. Second, based on the same evidence, it seems reasonable to bound $E[Y]$ from above by a value close to $E[Y|Z=1]$. To the best of our knowledge, there is no instance in which the evidence suggests that the non-participating population would score substantially higher than the participating population if they participated in an assessment. Therefore, we begin with the conservative, but reasonable, strategy to use $Q_{60}(Y|Z=1)$ as an upper bound on $E[Y|Z=0]$. Accordingly, in Figure 1 we report interval estimates under assumption A1.1 with $\alpha_L = 0.25$ and $\alpha_U = 0.60$, for each education system participating in ICILS 2018. Additionally, in Figure 1 we also report point estimates under ignorability assumptions A2 and A3, which correspond to the standard reporting strategy implemented in ICILS and many other ILSA studies. Table A1 in the Appendix provides all necessary estimates to reconstruct these reported bounds.

**Figure 1 Reporting strategy under assumption A1.1 with $\alpha_L = 0.25$ & $\alpha_U = 0.60$**

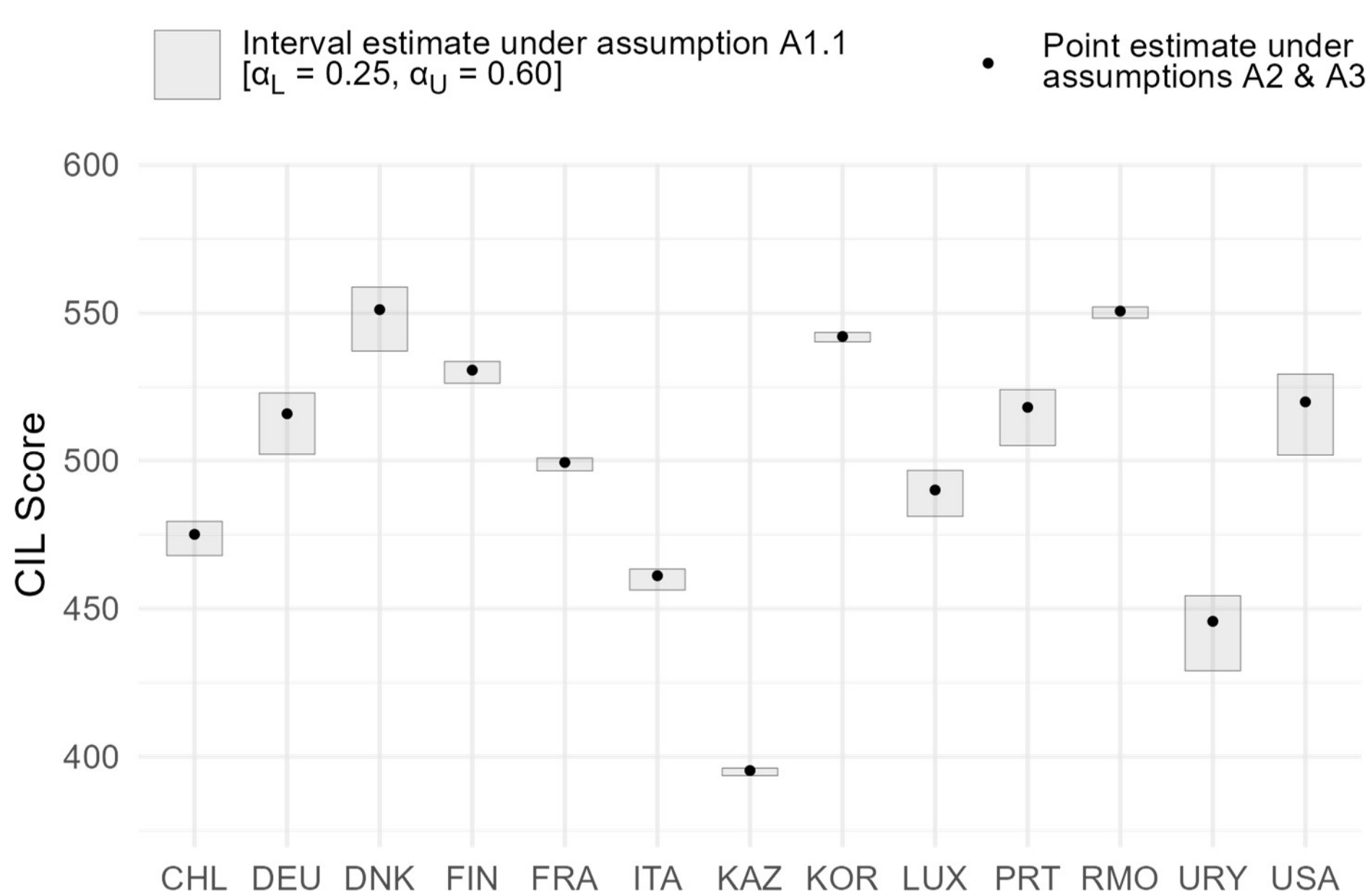

Figure 1 shows that, under assumption A1.1, estimation of $E[Y]$ is more ambiguous in education systems where the rate of student non-participation is relatively high. For example, with a non-participation rate of 39%, the USA has the highest rate across all education systems assessed by ICILS. The USA also has the widest interval estimate of $E[y]$—i.e., the most ambiguous results—with a length of about 27 score-points. Interestingly, Uruguay has the second-widest interval estimate, despite having a lower rate of student non-participation (27%) than Denmark (35%), Germany (31%), and Portugal (31%). Recall that the length of each interval estimate in Figure 1 is given by $\left(1-\hat{p}\right) * (\hat{q}_{60} - \hat{q}_{25})$. Therefore, the quantity $(\hat{q}_{60} - \hat{q}_{25})$ must be larger in Uruguay than in Denmark, Germany, and Portugal. That is, the estimated set of logically possible values that $E[Y|Z=0]$ can take under assumption A1.1 is larger in Uruguay than in these other countries, because this estimate of the spread of the achievement distribution among participating students is larger in Uruguay than in these other three nations.

We note here that Figure 1 reports estimates of $E[Y]$ under two alternative estimation strategies discussed in Section 2. First, interval estimation under assumption A1.1 is likely to be more credible—to the extent that it imposes weaker restrictions on $E[Y|Z=0]$—but also generates more ambiguous findings. On the other hand, point estimation under ignorability assumptions A2 and A3 is unambiguous but the most restrictive about $E[Y|Z=0]$ and therefore pushes to the limit the Law of Decreasing Credibility, which states that "the credibility of inference decreases with the strength of the assumptions maintained" (Manski, 2003, pp. 1).

It is also worth noting that, even under relatively weak assumptions on achievement among non-participants, notable differences arise across countries if one were to abstract from concerns about sampling imprecision. Of particular note are the findings for Denmark, which has the second-highest non-participation rate at 35% and is tied with the city of Moscow from the Russian Federation, which participated as a single entity in ICILS 2018, for the highest mean achievement score of 551 among participating students. Even with this high rate of non-participation, the estimated lower bound for Denmark of 537 exceeds the estimated upper bound on mean achievement for every education system other than the city of Moscow and Korea.

### 4.2 Reporting strategy under monotone non-participation

A major concern regarding the causes of student non-participation in ILSA studies is that there may be some systematic disincentives to participate among lower achieving schools and students, such that members of the non-participating population are, on average, less proficient than are participants. Given this concern, which is consistent with previously cited evidence from ILSA non-participation studies such as Brown et al. (2007), it is reasonable to bound $E[Y]$ from above by $E[Y|Z=1]$. In Figure 2, we present interval estimates of mean achievement in the ICILS assessment under the assumption that $Q_{25}(Y|Z=1) \leq E[Y|Z=0] \leq E[Y|Z=1]$. Thus, the lower bounds of the interval estimates presented in Figure 1 and Figure 2 are identical, but the upper bounds here are lower to the extent that the estimate of $E[Y|Z=1]$ is less than the estimate of $Q_{60}(Y|Z=1)$.

For each education system, the interval estimate presented in Figure 2 is less wide and, hence, generates less ambiguous findings on $E[Y]$. To gauge the informativeness of this monotonicity assumption, we note that the difference between the interval estimates presented in Figure 1 and those presented in Figure 2 is given by $\left(1-\hat{p}\right)(\hat{q}_{60} - \hat{\mu})$; that is, a function of the empirical distribution of student achievement and the rate of student non-participation. Table A2 in the Appendix provides all necessary estimates to reconstruct these reported bounds.

**Figure 2 Reporting strategy under monotone non-participation**

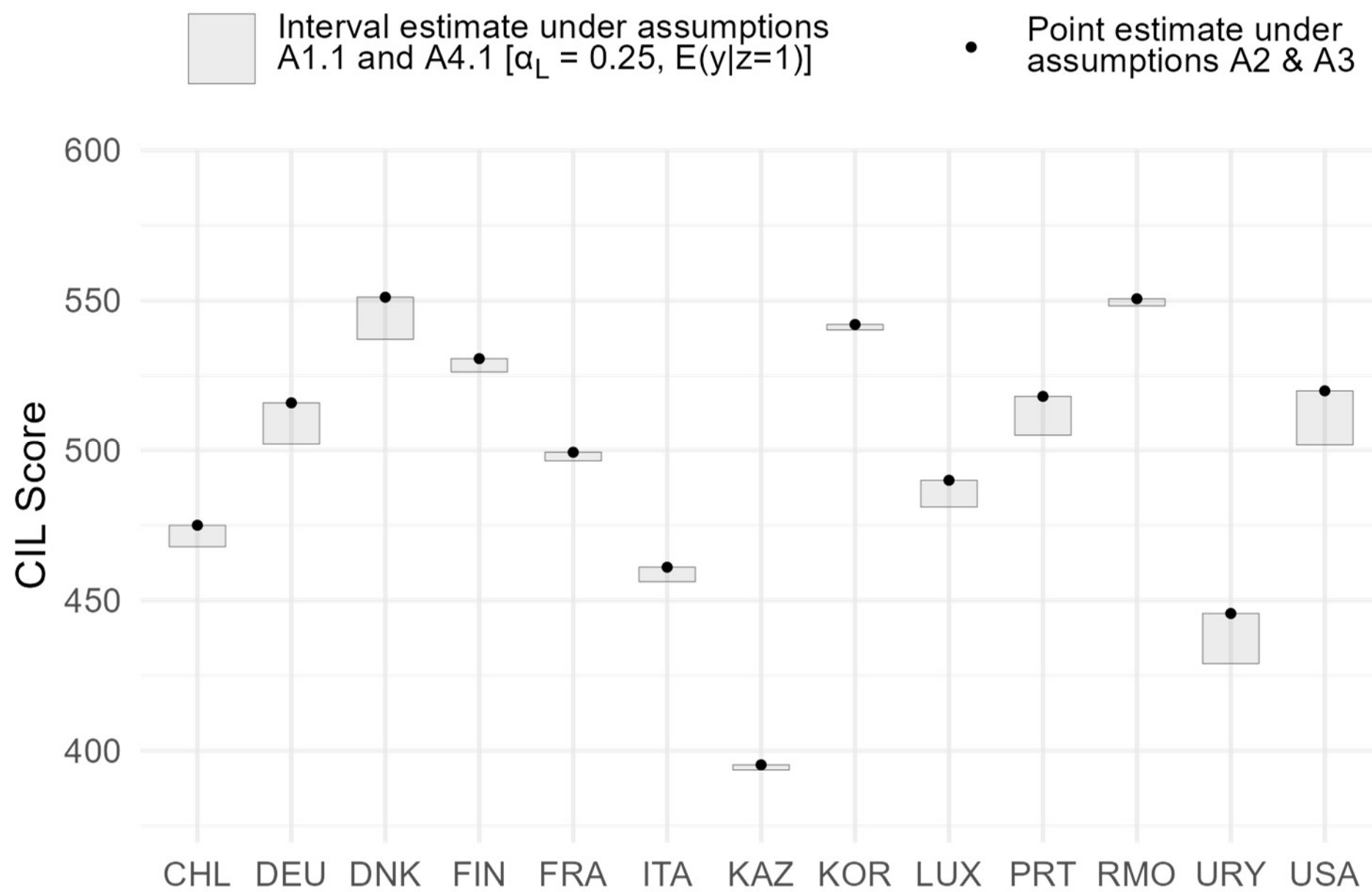

### 4.3 A hybrid reporting strategy: credible assumptions and standard practices

ILSA studies have traditionally reported point estimates under ignorability assumptions. However, recognizing that non-participation poses a fundamental threat to the validity of the reported findings, it is customary to include cautionary annotations in the reports if participation rates fall below a predetermined threshold. In the context of ICILS and other IEA studies, achieved participation rates are labeled as *Category 1*, *Category 2*, or *Category 3*. Using the notation introduced in Section 2.2, Category 1 implies that, using information collected solely from sampled schools, (i) the estimates of $P(Z_1 = 0)$ and $P(Z_1 = 1, Z_2 = 0)$ are each below 0.15, and (ii) the estimate of $P(Z_1 = 1, Z_2 = 1)$ is larger than 0.75. Similarly, Category 2 implies that the same thresholds were met after the inclusion of replacement schools; that is, acting as if recipient schools had never been sampled but instead their replacements had been selected for the original sample of schools. Finally, Category 3 implies that these thresholds were not met even after the inclusion of replacement schools. As argued in Section 2.2, the use of replacement schools has no obvious purpose in an identification analysis; therefore, we do not distinguish here between Category 2 and Category 3. See Tieck (2020b, p. 91) for a more detailed description of this categorization.

This reporting policy can be incorporated into our present analysis. For example, Dominitz and Manski (2025) note that the potential bias under ignorability assumptions is given by $P(Z = 0) \cdot \big[E[Y|Z = 1] - E[Y|Z = 0]\big]$. Therefore, the reporting of estimates that assume ignorability may be reasonable if either $P(Z = 0)$ or $\big[E[Y|Z = 1] - E[Y|Z = 0]\big]$ are known to be small. In the absence of knowledge about $\big[E[Y|Z = 1] - E[Y|Z = 0]\big]$—because $E[Y|Z = 0]$ is unobservable—one may be willing to base survey policy on $P(Z = 0)$ alone, as is the standard practice with ILSA studies. Based on this, and in the context of ICILS, one could choose to report point estimates under ignorability assumptions if participation rates in an education system fall into Category 1; otherwise, one would report interval estimates under more reasonable and credible assumptions, such as those presented in Figures 1 and 2. While we believe that a uniform approach towards the reporting of ambiguity is more appropriate, we recognize the potential appeal of such an intermediate step given the standard practices in the profession.

In Figure 3 we show how this reporting strategy could be implemented in the context of ICILS. Table A3 in the Appendix provides all necessary estimates to reconstruct these reported bounds.

**Figure 3 Reporting strategy under credible assumptions and standard survey policy**

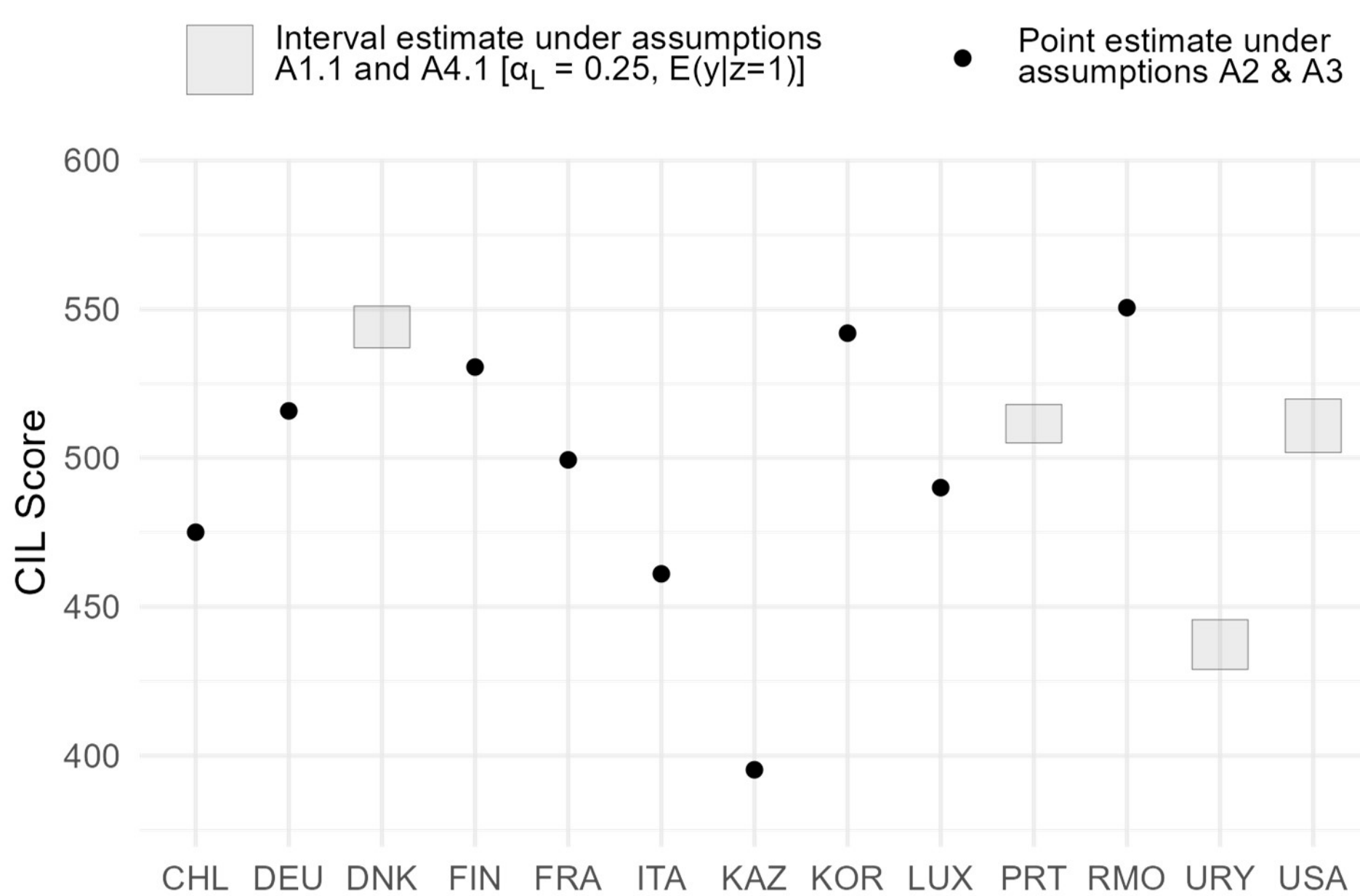

In ICILS 2018, Denmark, Portugal, Uruguay, and the USA did not meet the participation requirements to be placed in Category 1; therefore, the inferences reached under ignorability assumptions are reported with cautionary annotations. Our approach quantifies the implications of the cautionary annotations, such as for achievement in Denmark, which still ranks within the top three among the top three performing education systems based on the interval estimate reported in Figure 3.

With the methodology depicted in Figure 3, we argue that, rather than reporting estimates as if the ignorability assumption must be maintained but then annotated if not credible, one could instead simply report what is believed to be reasonable and credible even if it is, to some extent, ambiguous. Put differently, Figure 3 is a form of facing ambiguity in a transparent way and acknowledging that it is “far better to approximate an answer to the right question, which is often vague, than the exact answer to the wrong question, which can always be made precise” (Tukey, 1962, pp. 13).

## 4. Conclusions

In this paper, we perform a partial identification analysis of non-participation in the context of multi-stage stratified sample designs as typically implemented in ILSA studies. This methodological framework provides a useful strategy for reporting interval estimates of population parameters that explicitly reflect the ambiguity arising from non-participation (Manski, 2016). We also inspect the standard model of non-participation typically maintained in ILSA studies and place it within the partial identification framework. Finally, we use our methodological findings to conduct an international comparison of academic achievement in an ILSA study, the 2018 ICILS, where we abstract from concerns about sampling imprecision.

We see two possible and immediate extensions to this analysis. The first one would incorporate statistical inference into the estimation approach. Tamer (2010) provides a notable review of the literature on this topic, emphasizing the use of subsampling techniques to construct confidence regions of a partially identified parameter. He notes that this strand of research is divided into approaches that construct confidence regions for an identified set-valued parameter (see, for example, Chernozhukov et al. (2007)) and approaches that construct confidence regions for a "true" parameter even though it cannot be point identified (see, for example, Imbens and Manski (2004)). In the context of ILSA studies, we see an immediate application for the former approach as one could, for example, use re-sampling techniques to approximate the sampling variance of the estimated bounds derived in this article. Techniques such as the Jackknife Repeated Replication (Tukey, 1958) could be useful for this, as that is the standard approach employed in ILSA studies to account for the complex sample design with which the data are collected. This approach could account for the varying selection probabilities, the stratification design, and the clustering effect that arise from the multi-stage nature of the sampling and data collection process, all of which affect the sampling distribution of parameter estimates.

Second, we focus here on reporting interval estimates of population parameters. We recognize, however, that many will find such reporting insufficient, having become accustomed to seeing and using point estimates with an associated standard error, confidence interval, or so-called *margin of error*. Dominitz and Manski (2017) considered the problem of predicting the partially identified population mean from sample data where non-participation is not known to be ignorable. For point prediction, they proposed a midpoint predictor—in particular, the midpoint of the interval estimate—that has desirable properties. For instance, the maximum mean square error (MSE) of this predictor is smaller than that of the sample mean among participants. Further, if the interval were known rather than estimated, then the midpoint predictor would be the point predictor that minimizes maximum MSE among all point predictors.

Dominitz and Manski (2025) apply this approach to election polls, arguing that a midpoint estimate should be reported along with a *total margin of error* (TME) measure based on the maximum MSE of the midpoint predictor under maintained assumptions on non-sampling error, including non-participation in the poll. The TME accounts for both sampling variation and potential bias that may arise from non-participation. If non-participation is ignorable, then the TME reduces to the usual margin of error. The same methodology could be applied to the results of ILSA studies, which would then satisfy desires for the reporting of point estimates along with a measure that characterizes sampling error.

## Acknowledgements

We gratefully acknowledge funding and support from the IEA Research and Development Funds. We are grateful to Chuck Manski, Mojca Rozman, two anonymous referees, and the editors of this journal for useful comments and suggestions. This research project was presented at the GEBF conference 2023 in Duisburg, Germany; at the IEA IRC in Dublin, Ireland; at the IEA General Assembly Meeting in Versailles, France; and, at the WERA conference 2023 in Singapore.